\documentstyle[12pt]{article}

\def\hybrid{\topmargin -20pt	\oddsidemargin 0pt
	\headheight 0pt	\headsep 0pt
	\textwidth 6.25in	
	\textheight 9.5in	
	\marginparwidth .875in
	\parskip 5pt plus 1pt	\jot = 1.5ex}

\hybrid

\def\baselinestretch{1.2}

\catcode`\@=11

\def\marginnote#1{}
\newcount\hour
\newcount\minute
\newtoks\amorpm
\hour=\time\divide\hour by60
\minute=\time{\multiply\hour by60 \global\advance\minute by-\hour}
\edef\standardtime{{\ifnum\hour<12 \global\amorpm={am}%
	\else\global\amorpm={pm}\advance\hour by-12 \fi
	\ifnum\hour=0 \hour=12 \fi
	\number\hour:\ifnum\minute<10 0\fi\number\minute\the\amorpm}}
\edef\militarytime{\number\hour:\ifnum\minute<10 0\fi\number\minute}

\def\draftlabel#1{{\@bsphack\if@filesw {\let\thepage\relax
   \xdef\@gtempa{\write\@auxout{\string
      \newlabel{#1}{{\@currentlabel}{\thepage}}}}}\@gtempa
   \if@nobreak \ifvmode\nobreak\fi\fi\fi\@esphack}
	\gdef\@eqnlabel{#1}}
\def\@eqnlabel{}
\def\@vacuum{}
\def\draftmarginnote#1{\marginpar{\raggedright\scriptsize\tt#1}}

\def\draft{\oddsidemargin -.5truein
	\def\@oddfoot{\sl preliminary draft \hfil
	\rm\thepage\hfil\sl\today\quad\militarytime}
	\let\@evenfoot\@oddfoot	\overfullrule 3pt
	\let\label=\draftlabel
	\let\marginnote=\draftmarginnote
   \def\@eqnnum{(\theequation)\rlap{\kern\marginparsep\tt\@eqnlabel}%
\global\let\@eqnlabel\@vacuum}  }

\def\preprint{\twocolumn\sloppy\flushbottom\parindent 2em
	\leftmargini 2em\leftmarginv .5em\leftmarginvi .5em
	\oddsidemargin -.5in	\evensidemargin -.5in
	\columnsep .4in	\footheight 0pt
	\textwidth 10.in	\topmargin  -.4in
	\headheight 12pt \topskip .4in
	\textheight 6.9in \footskip 0pt
	\def\@oddhead{\thepage\hfil\addtocounter{page}{1}\thepage}
	\let\@evenhead\@oddhead	\def\@oddfoot{}	\def\@evenfoot{} }

\def\numberbysection{\@addtoreset{equation}{section}
	\def\theequation{\thesection.\arabic{equation}}}

\def\underline#1{\relax\ifmmode\@@underline#1\else
	$\@@underline{\hbox{#1}}$\relax\fi}

\def\titlepage{\@restonecolfalse\if@twocolumn\@restonecoltrue\onecolumn
     \else \newpage \fi \thispagestyle{empty}\c@page\z@
	\def\thefootnote{\fnsymbol{footnote}} }

\def\endtitlepage{\if@restonecol\twocolumn \else \newpage \fi
	\def\thefootnote{\arabic{footnote}}
	\setcounter{footnote}{0}}  

\catcode`@=12
\relax

\def\figcap{\section*{Figure Captions\markboth
	{FIGURECAPTIONS}{FIGURECAPTIONS}}\list
	{Figure \arabic{enumi}:\hfill}{\settowidth\labelwidth{Figure 999:}
	\leftmargin\labelwidth
	\advance\leftmargin\labelsep\usecounter{enumi}}}
 \relax
\def\tablecap{\section*{Table Captions\markboth
	{TABLECAPTIONS}{TABLECAPTIONS}}\list
	{Table \arabic{enumi}:\hfill}{\settowidth\labelwidth{Table 999:}
	\leftmargin\labelwidth
	\advance\leftmargin\labelsep\usecounter{enumi}}}
 \relax
\def\reflist{\section*{References\markboth
	{REFLIST}{REFLIST}}\list
	{[\arabic{enumi}]\hfill}{\settowidth\labelwidth{[999]}
	\leftmargin\labelwidth
	\advance\leftmargin\labelsep\usecounter{enumi}}}
 \relax
\makeatletter
\newcounter{pubctr}
\def\publist{\@ifnextchar[{\@publist}{\@@publist}}
\def\@publist[#1]{\list
	{[\arabic{pubctr}]\hfill}{\settowidth\labelwidth{[999]}
	\leftmargin\labelwidth
	\advance\leftmargin\labelsep
	\@nmbrlisttrue\def\@listctr{pubctr}
	\setcounter{pubctr}{#1}\addtocounter{pubctr}{-1}}}
\def\@@publist{\list
	{[\arabic{pubctr}]\hfill}{\settowidth\labelwidth{[999]}
	\leftmargin\labelwidth
	\advance\leftmargin\labelsep
	\@nmbrlisttrue\def\@listctr{pubctr}}}
 \relax
\makeatother
\newskip\humongous \humongous=0pt plus 1000pt minus 1000pt

\newif\ifdtup

\relax

\def\thefootnote{\fnsymbol{footnote}}
\def\be{\begin{equation}}
\def\ee{\end{equation}}
\def\ba{\begin{eqnarray}}
\def\ea{\end{eqnarray}}

\begin{document}
\renewcommand{\theequation}{\arabic{equation}}
\newcommand{\beq}{\begin{equation}}
\newcommand{\eeq}[1]{\label{#1}\end{equation}}
\newcommand{\ber}{\begin{eqnarray}}
\newcommand{\eer}[1]{\label{#1}\end{eqnarray}}
\begin{titlepage}
\begin{center}

\hfill CERN-TH/96-148\\
\hfill hep-th/9606030\\

\vskip .5in

{\large \bf 2-D GRAVISOLITONS IN STRING THEORY}
\footnote{Contribution to the proceedings of the 2nd 
International Sakharov Conference, 20--24 May 1996, Moscow,
Russia; to be published by World Scientific}\\ 

\vskip 0.7in

{\bf Ioannis Bakas}
\footnote{Permanent address: Department of Physics, University 
of Patras, GR-26110 Patras, Greece}
\footnote{e-mail address: BAKAS@SURYA11.CERN.CH, 
BAKAS@NXTH04.CERN.CH}\\
\vskip .1in

{\em Theory Division, CERN\\  
     CH-1211 Geneva 23, Switzerland}\\

\vskip .8in

\end{center}

\vskip .4in

\begin{center} {\bf ABSTRACT } \end{center}
\begin{quotation}\noindent
Several gravitational string backgrounds can be 
interpreted as 2-dim soliton solutions of 
reduced axion-dilaton gravity. They include 
black-hole and worm-hole solutions as well as
cosmological models with an exact conformal 
field theory description. We illustrate the
use of gravisolitons for the particular 
example of Nappi-Witten universe which is 
thus ``created" from flat space by soliton 
dressing. We also make some general comments 
about the status of gravisolitons in comparison
to soliton solutions of other 2-dim integrable 
systems without gravity.
\end{quotation}
\vskip1.0cm
CERN-TH/96-148 \\
June 1996\\
\end{titlepage}
\vfill
\eject
\def\baselinestretch{1.2}
\baselineskip 16 pt
\noindent
Several years ago it was realized that the sector 
of gravitational theories obtained by dimensional 
reduction to two space-time 
dimensions is integrable. Then, the 
hidden symmetries of the resulting 2-dim non-linear
equations were successfully employed as solution 
generating transformations to construct
new backgrounds from old ones. This was originally
done for pure gravity and it was subsequently generalized
to include Maxwell fields as well as other 
matter fields that arise in various reduced supergravity
models. The classical problem is exactly solvable in the
sense that the hidden symmetry group of the equations
is an infinite dimensional current group 
($\widehat{SU}(1,1)$ for pure gravity, $\widehat{SU}(2,1)$ 
for Einstein-Maxwell, 
and so on) and any solution can be obtained from any 
other one by transitive action. The integration can be 
performed by applying the techniques of inverse 
scattering as it was first proposed by Belinski and 
Sakharov [1]. It requires the introduction of an associated
Lax-pair that linearizes the reduced system of equations 
and it makes use of a spectral parameter. 
This prescription is of course rather formal and only in
certain cases that involve transformations with first
order poles in the (complex) spectral parameter plane 
the calculations can be performed explicitly. Group
elements with such a special pole 
structure are called soliton 
dressing transformations and the resulting gravitational 
backgrounds are the (multi)-soliton excitations of a 
given seed solution. From the mathematical point of view
these excitations have the full status of solitons. 
As we will see later, however, 
the 2-dim gravisolitons also have some differences 
from the usual soliton solutions arising in other 2-dim 
integrable systems without gravity. 

We are interested in investigating the precise meaning and 
use of 2-dim gravisolitons because they arise in the
sector of string theory where the compactification goes 
down to two space-time dimensions. Discrete remnants of the 
corresponding infinite dimensional hidden symmetry group
may naturally qualify as U-duality symmetries, and hence this 
particular sector of strings is expected to be quite
rich in symmetry and probably exactly solvable by 
algebraic methods. Since solitonic configurations play
a crucial role in this modern line of investigation 
in string theory and the soliton spectrum usually becomes 
bigger the lower we go in dimension, we are faced with
the notion of 2-dim gravisolitons in this sector, hoping to 
make further progress by understanding them better. Any 
such configuration has an interpretation as solution of the
original higher dimensional theory with (commuting) 
Killing symmetries, but it will not necessarily be
quantum mechanically stable in higher dimensions. Even in 
two (reduced) dimensions the role of gravisolitons is
not well understood and their relevance still remains 
somewhat obscure. For example, one would like to 
establish a connection between their conventional 
formulation as inverse scattering solitons and their 
properties under the 2-dim reduction of the space-time 
supersymmetry algebra. It is well known that in other
(non-gravitational) integrable systems in two dimensions, 
like the sine-Gordon model, the solitons saturate the 
Bogomolny bound of the rigid supersymmetry algebra 
and there is the notion of their topological charge 
that becomes important for supersymmetry charges. The theory 
of gravisolitons is at a more primitive and formal level 
at this stage. 

It is interesting that many solutions of Einstein equations
can be described as 2-dim gravisolitons including the
most general Kerr-NUT black-holes in four dimensions 
(exploiting their stationary axisymmetric character) as 
well as various cosmological and colliding plane wave 
solutions. Recently we have extended this framework to
describe some gravitational string backgrounds with 
non-trivial dilaton and axion fields as solitons on flat
space [2]. Here we will only examine the details of the 
construction for cosmological backgrounds with an application
to the Nappi-Witten universe which is known to correspond 
to the exact conformal field theory coset 
$SL(2) \times SU(2) / SO(1,1) \times U(1)$ [3]. 
In the present context we consider 4-dim metrics
with two commuting Killing symmetries
\be
ds^2 = f(X^0 , X^1) (-(dX^0)^2 + (dX^1)^2) + 
g_{AB} (X^0 , X^1) dX^A dX^B ~; ~~~A, B = 2, 3
\ee
together with axion $b(X^0 , X^1)$ and dilaton 
$\Phi (X^0 , X^1)$, which in the lowest order effective
theory couple to 4-dim gravity as an $SL(2)/R$ 
$\sigma$-model with matrix parametrization
\be
\lambda = e^{2 \Phi} \left(\begin{array}{ccc}
1 & & b\\
  & &  \\
b & & b^2 + e^{-4 \Phi} \end{array} \right).
\ee
All this is directly formulated in the Einstein frame
which is related to the $\sigma$-model frame of the string
metric by $G_{\mu \nu}^{(\sigma)} = e^{2 \Phi} G_{\mu \nu}$.
The cosmological solution of Nappi-Witten falls into this 
class having
\ba
f & = & e^{-2 \Phi} = 1 - \cos 2X^0 \cos 2X^1 + \sin \theta 
(\cos 2X^0 - \cos 2X^1) , \\
g_{AB} & = & \left(\begin{array}{ccc}
{\sin}^2 X^0 {\sin}^2 X^1 & & 0\\
      &   &   \\
0 &   & {\cos}^2 X^0 {\cos}^2 X^1 \end{array} \right) ,\\
b & = & \cos \theta (\cos 2X^0 - \cos 2X^1)
\ea
and it provides a model of an expanding and recollapsing 
universe as $X^0$ (as well as $X^1$) range from 0 to $\pi/2$.
The parameter $\theta$ is free depending on the gauging of 
the underlying conformal coset model. 

The 2-dim system that arises after dimensional reduction of 
the lowest order effective theory is given in general by two
decoupled Ernst-type $SL(2)/R$ $\sigma$-models, one for the 
metric sector and one for the axion-dilaton sector, 
\be
{\partial}_{+}(\sqrt{\det g} g^{-1} {\partial}_{-} g) + 
{\partial}_{-}(\sqrt{\det g} g^{-1} {\partial}_{+} g) = 0, 
\ee
\be
{\partial}_{+}(\sqrt{\det g} {\lambda}^{-1} 
{\partial}_{-} \lambda) + {\partial}_{-}(\sqrt{\det g} 
{\lambda}^{-1} {\partial}_{+} \lambda) = 0,
\ee
where we have introduced for convenience the light-cone 
variables $(X^0 \pm X^1)/2$. The same equation is satisfied
by $\sqrt{\det g} \lambda$, thus putting both sectors 
on exactly the same footing. Also note that since 
${\partial}_{+} {\partial}_{-} \sqrt{\det g} = 0$ we may
always choose without loss of generality (and from now on)
$X^0 = \sqrt{\det g} := \alpha$
and call $\beta$ the corresponding conjugate solution. 
Then, the equations for the conformal factor simply become
\be
{\partial}_{\pm} (\log f) = -{1 \over \alpha} + {\alpha \over 4}
{\rm Tr} \left( (g^{-1} {\partial}_{\pm} g)^2 + 
({\lambda}^{-1} {\partial}_{\pm} \lambda)^2 \right)
\ee
and they can be integrated by quadratures. 
The integrability of the lowest order effective theory on 
these backgrounds clearly 
originates from the integrability of the Ernst $\sigma$-models. 

Next we examine the structure of the soliton solutions by 
applying inverse scattering methods. Introducing a spectral
parameter $l$ we consider the linear system
\be
\left( {\partial}_{\pm} \mp {2l \over l \mp \alpha} {\partial}_{l} 
\pm {\alpha \over l \mp \alpha} 
({\partial}_{\pm} g) g^{-1} \right) \Psi = 0,
\ee
where $\Psi (l = 0) = g$. The compatibility condition is precisely
the Ernst equation (6); similarly we could have replaced $g$
with $\alpha \lambda$ to apply the same method to (7). 
Let $g_{0}$ be a seed solution with ${\Psi}_{0}$ being the 
corresponding solution of (9). The $n$-soliton excitation has
\be
\Psi (l) = \left( 1 + \sum_{k=1}^{n} {R_{k} \over 
l - {\mu}_{k}} \right) {\Psi}_{0} (l),
\ee
where the poles are roots of the algebraic equation
\be
{\mu}_{k}^2 + 2(\beta - C_{0}^{(k)}) {\mu}_{k} + 
{\alpha}^2 = 0
\ee
for arbitrary constants $C_{0}^{(k)}$. The residues are 
$2 \times 2$ degenerate matrices. Here, in view of the 
following application, we only give 
their explicit form for $n=1$, suppressing the soliton index, 
\be
R_{AB} = \left(\mu - {{\alpha}^2 \over \mu} \right) 
{\sum_{C} M_{C} (g_{0})_{AC} M_{B} \over \sum_{C,D} 
M_{C} (g_{0})_{CD} M_{D}}, 
\ee
where the 2-component vector $M$ is
$M_{B} = \sum_{A} C_{A} {\Psi}_{0}^{-1} (l = \mu)_{AB}$
with arbitrary moduli parameter vector $C = (C_{1} , C_{2})$. 
Introducing appropriate normalization so that $\sqrt{\det g}$
is preserved under the soliton dressing we find the general 
1-soliton excitation of $g_{0}$,
\be
g = {\mu \over \alpha} \left( 1 - {R \over \mu} \right) g_{0}.
\ee

The Nappi-Witten universe arises as an $(1, 1)$-solitonic 
excitation, where the two soliton numbers 
are referring to the metric
and axion-dilaton sectors respectively. For this we use as 
seed string background the following solution (in the Einstein
frame):
\be
ds^2 = - {d \alpha}^2 + {d \beta}^2 + \alpha 
(dz^2 + dw^2) ; ~~~~~ 
b_{0} = 0 , ~~~~ e^{-2 {\Phi}_{0}} = \alpha ,  
\ee
which is a T-dual face of flat space with trivial axion 
and dilaton fields with respect to the isometry 
$\partial / \partial \beta$. Applying the previous construction
to the $g$ and $\lambda$ sectors of the theory and choosing 
respectively for the moduli parameters
\be
C_{0} = -1, ~~~~ C_{1} = 0 ; ~~~~~
C_{0}^{\prime} = 1, ~~~~ {C_{2}^{\prime} \over 
2 C_{1}^{\prime}} = {\sin \theta - 1 \over \cos \theta}
\ee
we arrive at the result; further details can be found 
in [2]. We 
only note here that the identification with the cosmological
background (3)--(5) is better described by introducing the
coordinate parametrization 
$\alpha = \sin 2X^0 \sin 2X^1$,  
$\beta = \cos 2X^0 \cos 2X^1$
and then the soliton moduli parameter $\theta$ corresponds 
to the arbitrariness in gauging the underlying conformal field 
theory coset. Thus, the expanding and recollapsing universe
of Nappi-Witten is actually a simple solitonic excitation of
the trivial string background. A similar interpretation can
be given to other conformal field theory backgrounds, but
lack of space does not permit to include more examples.

Finally we briefly comment on the present status of 
2-dim soliton solutions in gravitational theories in 
connection with soliton solutions of other integrable 
systems. From (11) we already see a difference from other
integrable systems in that the soliton poles in the 
spectral parameter space have their own space-time 
dynamics
${\partial}_{\pm} {\mu}_{k} = 2 {\mu}_{k} /  
\alpha \mp {\mu}_{k}$. 
Disentangling the soliton interaction from the background 
geometry and looking for physically meaningful quantities
that usually characterize solitons in rigid theories is 
not a simple matter because of the transformation of the 
conformal factor $f$ in (8). 
It was suggested recently that the
integer quantity ${\rm sgn}({\alpha}^2 - {\mu}^2)$ plays 
the role of a topological index for gravisolitons [4]. 
Note that the two solutions of (11), ${\mu}_{+}$ and 
${\mu}_{-} = {\alpha}^2 / {\mu}_{+}$, can be regarded
as defining a soliton or anti-soliton configuration with
${\alpha}^2 - {\mu}^2$ changing sign under
${\mu}_{+} \leftrightarrow {\mu}_{-}$. It is essential for 
all this to choose globally ${\mu}_{+}$ or ${\mu}_{-}$ even
through the causally disconnected regions of 
space-time. It can
be seen in the axion-dilaton sector of the Nappi-Witten 
universe that under $\theta \rightarrow \theta + \pi$ the
soliton interpretation of the background changes to 
that of an anti-soliton.

It will be very interesting for future study to examine
the status of gravisolitons using the 2-dim reduction of 
the space-time supersymmetry algebra. The conformal factor
will play a crucial role in this context as it describes 
central extensions of the underlying hidden symmetry group, 
where the supersymmetry transformations can be embedded 
by bosonization [5]. The physics of 2-dim gravisolitons 
remains unexplored to a large extend.

\vskip1.0cm
\centerline{\bf References}
\begin{enumerate}
\item V. Belinski and V. Sakharov, Sov. Phys. JETP 
\underline{48} (1978) 985; ibid \underline{50} (1979) 1.
\item I. Bakas, ``Solitons of axion-dilaton gravity", 
CERN-TH/96-121, hep-th/9605043.
\item C. Nappi and E. Witten, Phys. Lett. \underline{B293} 
(1992) 309.
\item V. Belinski, Phys. Rev. \underline{D44} (1991) 3109.
\item H. Nicolai and N. Warner, Comm. Math. Phys. 
\underline{125} (1989) 369.
\end{enumerate}
\end{document}